\documentstyle[11pt]{article}
\begin{document}
\baselineskip 18pt
\begin{titlepage}
\centerline{\large\bf  New Viewpoint to the Source of Weak CP Phase}
\vspace{1cm}

\centerline{Jing-Ling Chen$^1$, Mo-Lin Ge$^{1,2}$,
            Xue-Qian Li$^{3,4}$, Yong Liu$^1$}
\vspace{0.5cm}
{\small
\centerline{\bf 1. Theoretical Physics Division}
\centerline{\bf Nankai Institute of Mathematics}
\centerline{\bf Nankai University, Tianjin 300071, P.R.China}
\vspace{0.2cm}
\centerline{\bf 2. Center for Advanced Study }
\centerline{\bf Tsinghua University, Beijing 100084, P.R.China}
\vspace{0.2cm}
\centerline{\bf 3. CCAST(World Laboratory) P.O.Box 8730, Beijing 100080,
                P.R.China}
\vspace{0.2cm}
\centerline{\bf 4. Department of Physics }
\centerline{\bf Nankai University, Tianjin 300071, P.R.China}
}
\vspace{2cm}

\centerline{\bf Abstract}
\vspace{0.3cm}

The relation between CP-violation phase angle and the other three
mixing angles in Cabibbo-Kobayashi-Maskawa matrix is postulated.
This relation has a very definite geometry meaning. The numerical
result coincides surprisingly with that extracting from the experiments.
It can be further put to the more precise tests in the future.\\ \\
PACS number(s): 11.30.Er, 12.10.Ck, 13.25.+m

\end{titlepage}

\centerline{\large\bf  New Viewpoint to the Source of Weak CP Phase}

\vspace{1cm}

Although more than thirty years have elapsed since the discovery of CP
violation [1], our understanding about the source of CP violation
is still very poor. In the Minimal Standard Model (MSM), CP violation
is due to the presence of a weak phase in the Cabibbo-Kabayashi-Maskawa
(CKM) matrix [2][3]. It is generally believed to be independent of the three
mixing angles.

Up to now, all the experimental results are in good agreement with MSM.
Nevertheless, the correctness of CKM mechanism is far from being proved.
On other hand, the MSM which possesses many free parameters is not fully
satisfactory.
People have tried to reduce the number of the free parameters, but
searching for the source of CP violation is more profound in
high energy physics [4]
[5][6][7][8]. Fritzsch
[9][10] noticed
that because the eigenstates of the weak interaction are not the quark
mass-eigenstates, there should be a unitary transformation to connect the
two bases. It would establish a certain relation between the
quark masses and the weak interaction mixing angles, while a weak CP
phase is embedded explicitly.

From the general theory of Kabayashi-Maskawa
[2],
we know that there can exist a phase factor in the three-generation
CKM matrix and it cannot be rotated away by re-defining the phases of
quarks, but we can ask whether there is an intrinsic relation between
the phase and the three rotation angles.

Concretely, supposing $V_d$ and $V_u$ diagonalize the mass matrices for
d-type and u-type quarks respectively [11], $V_{KM}
\equiv V_u^{\dagger}V_d$
is the CKM matrix and can be written as
\begin{equation}
V_{KM}= \left (
\begin{array}{ccc}
   c_1 & -s_1c_3& -s_1s_3 \\
   s_1c_2 & c_1c_2c_3-s_2s_3e^{i\delta}& c_1c_2s_3+s_2c_3e^{i\delta}\\
   s_1s_2 & c_1s_2c_3+c_2s_3e^{i\delta}& c_1s_2s_3-c_2c_3e^{i\delta}
\end{array}
\right )
\end{equation}
with the standard notations $s_i=\sin\theta_i$ and $c_i=\cos\theta_i$.

It is noted that here we adopt the original form of the CKM parametrization.
There are some other parametrization ways, for example the Wolfenstein's
[12][13][14]
and that recommended by the data group [15][16]
[17], but it is believed that physics
does not change when adopting various parametrizations.

It is well known that the KM parametrization can be viewed as a product of
three Eulerian rotation matrices and a phase matrix [11]
\begin{equation}
V_{KM}= \left (
\begin{array}{ccc}
   1 & 0 & 0 \\
   0& c_2 &-s_2 \\
   0 & s_2 & c_2
\end{array}
\right )
\left (
\begin{array}{ccc}
   c_1 &-s_1 & 0 \\
   s_1& c_1 & 0 \\
   0 & 0 & 1
\end{array}
\right )
\left (
\begin{array}{ccc}
   1 &0& 0 \\
   0& 1& 0 \\
   0 & 0 & -e^{i\delta}
\end{array}
\right )
\left (
\begin{array}{ccc}
   1 &0& 0 \\
   0& c_3& s_3 \\
   0 & -s_3 & c_3
\end{array}
\right ).
\end{equation}

People have noticed that the weak CP phase $\delta$, which cannot
be eliminated in the three generation CKM matrix by any means, is introduced
artificially and seems to have nothing to do with the three "rotation"
angles. Anyway, such a fact does not seem to be natural.

However, for a naive $O(3)$ rotation group, a geometric phase can
automatically arise while two non-uniaxial successive rotation
transformations being performed. For
instance, $R_x(\theta_1)$ denotes a clockwise rotation about the x-axis
by $\theta_1$, while $R_y(\theta_2)$ is about the y-axis by $\theta_2$.
Supposing on a unit-sphere surface, the positive z-axis intersects with
the surface at a point A, after performing these two sequential operations
$R_y(\theta_2)R_x(\theta_1)$, the point-A would reach point-B via an
intermediate point-C, by contrast, one can connect A and B by a single
rotation $R_{\hat \xi}(\theta_3)$, where $R_{\hat \xi}(\theta_3)$ denotes a
clockwise rotation about the ${\hat \xi}-$axis by $\theta_3$. The geometric
meaning can be depicted in a more obvious way is that if one chooses an
arbitrary tangent vector
$\hat \alpha$ at point-A which would rotate to $\hat \alpha'$ and $\hat
\alpha \prime \prime$ by $R_y(\theta_2)
R_x(\theta_1)$ and $R_{\hat \xi}(\theta_3)$ respectively, then one can find
that $\hat \alpha'$ does not coincide with $\hat \alpha \prime \prime$, but
deviates by an extra rotation. Concretely, if one writes down the rotation
in the adjoint representation of $O(3)$, he can find
\begin{equation}
R_{\hat \eta}(\delta)R_{\hat \xi}(\theta_3)=R_y(\theta_2)R_x(\theta_1),
\end{equation}
where $R_{\hat \eta}(\delta)$ represents a counterclockwise rotation about
the ${\hat \eta}-$axis by $\delta$. The ${\hat \eta}-$axis is a vector from
the center of the sphere to the point-B and $\delta$ satisfies a relation
\begin{equation}
\cos{\delta\over 2}={1+\cos\theta_1+\cos\theta_2+\cos\theta_3\over
4\cos{\theta_1\over 2}\cos{\theta_2\over 2}\cos{\theta_3\over 2}}.
\end{equation}

Here $\delta$ is fully determined by the three rotation angles.
Geometrically, it is the solid angle which $\theta_1,\;\theta_2$ and
$\theta_3$ span, or in other words, this solid angle corresponds to
an area of the spherical triangle constructed by $\theta_i\;(i=1,2,3)$,
or the excess of the spherical triangle.

Eq.(3) can be recast in the form
\begin{equation}
R_{\hat \eta}(\delta)=R_y(\theta_2)R_x(\theta_1)R^{-1}_{\hat \xi}(\theta_3)
\end{equation}
to emphasize that the product of two finite rotations about different
axes cannot be cancelled out by a third rotation but leads to a residual
phase described by $R_{\hat \eta} (\delta)$. The appearance
of this phase is due to the non-commutativity of finite three-dimensional
rotations.
The same conclusion can be drawn for $ SU(2),\; SU(3)$ group and the Lorentz
group $O(3,1)$ etc.[18][19][20][21][22].
It is
worthy mentioning that the Wigner angle can be interpreted as a geometric
phase (or anholonomy) associated with a triangular circuit in rapidity
space in the theory of special relativity [23][24].

When a group is an Abelian one, the phase factor will not emerge.
Indeed, for the two-generation quarks, the mixing corresponds to a planar
rotation, all operations are commutative, there dose not
exist such an extra factor, correspondingly, we know that there is no a CP
phase if there were only two generations. This stimulates us to reach an
understanding
that for the three-generation case, the weak CP phase which cannot be
eliminated arises from a hidden $O(3)$ rotation symmetry in the flavor
space. Accepting this understanding,
the value of the mysterious weak phase $\delta$ will be determined
by the simple
relation Eq.(4) and has a definite
geometric meaning: it is just the solid angle
spaned by the three mixing angles $\theta_1,\;\theta_2$ and $\theta_3$ in
the flavor space. In fact, it has been recognized that the CP violation
parameter $\epsilon$ is related to a certain area [4][25]
more than ten years before, but the geometric relation between this area
and the three mixing angles and CP violation phase has not been recognized
yet.

So far the only reliably measured CP violation quantity is $\epsilon$ in
the K-system and the mechanism causing $K^0-\bar K^0$ mixing has already
been well studied in the framework of MSM. Except an unknown B-factor,
one can evaluate $\epsilon$ in terms of the CP phase $\delta$ as
[26][27][28]
\begin{equation}
|\epsilon|\approx \cos\theta_2\sin\theta_2\sin\theta_3\sin\delta
\left[{\sin^2\theta_2(1+\eta\log\eta)-\cos^2\theta_2\eta(1+\log\eta)
\over\sin^4\theta_2+\cos^4\theta_2\eta-2\sin^2\theta_2\cos^2\theta_2\eta\log
\eta}\right],
\end{equation}
where $\eta=m_c^2/m_t^2$.

The numerical results is listed in $Table \; 1$.
\begin{table}[t]
\caption{Comparison of Theoretical and Experimental Datum }
\begin{center}
\begin{tabular}{c|c|c|c|c}
\hline\hline
{\sl $V_{ud}$} & {\sl $V_{cd}$} & {\sl $V_{us}$} &
{\sl $sin\delta_{Th}$}& {\sl $sin\delta_{Ex}$} \\ \hline
0.9745& 0.218& 0.219& 0.022638& 0.002658\\ \hline
0.9748& 0.218& 0.219& 0.018978& 0.002746\\ \hline
0.9751& 0.218& 0.219& 0.014477& 0.002939\\ \hline
0.9754& 0.218& 0.219& 0.007791& 0.003524\\ \hline
0.9745& 0.220& 0.219& 0.019884& 0.002248\\ \hline
0.9748& 0.220& 0.219& 0.015635& 0.002255\\ \hline
0.9751& 0.220& 0.219& 0.009757& 0.002416\\ \hline
0.9745& 0.222& 0.219& 0.015439& 0.001830\\ \hline
0.9748& 0.222& 0.219& 0.009410& 0.002100\\ \hline
0.9745& 0.224& 0.219& 0.006480& 0.003807\\ \hline
0.9745& 0.218& 0.221& 0.018836& 0.003345\\ \hline
0.9748& 0.218& 0.221& 0.014269& 0.003837\\ \hline
0.9751& 0.218& 0.221& 0.007357& 0.005573\\ \hline
0.9745& 0.220& 0.221& 0.016697& 0.002829\\ \hline
0.9748& 0.220& 0.221& 0.011367& 0.003151\\ \hline
0.9745& 0.222& 0.221& 0.012755& 0.002302\\ \hline
0.9748& 0.222& 0.221& 0.003806& 0.002934\\ \hline
0.9745& 0.224& 0.221& 0.002658& 0.004790\\ \hline
0.9745& 0.218& 0.223& 0.012551& 0.005214\\ \hline
0.9748& 0.218& 0.223& 0.002690& 0.019451\\ \hline
0.9745& 0.220& 0.223& 0.011066& 0.004410\\ \hline
0.9745& 0.222& 0.223& 0.006865& 0.003589\\ \hline\hline
$Average$&$Values:$& & 0.0119$\pm$0.0056&0.0040$\pm$0.0035\\ \hline\hline
\end{tabular}
\end{center}
\end{table}
Where the inputs of $\mid V_{ij} \mid $ are taken from the data book
[29] and
$$m_c=1.5 \;{\rm GeV},\;\;\; m_t=176\; {\rm GeV},\;\;\; |\epsilon|=
2.3\times 10^{-3},$$
with all the given errors.
Here, $sin \delta_{Th}$ is calculated by using Eq.(4),
while $sin \delta_{Ex}$ extracted from Eq.(6).

One can notice that considering the experimental error tolerance,
the two obtained values are consistent. Since the
extraction of $\sin\delta$ from
the data of $\epsilon$ still depends on the evaluation of the concerned
hadronic
matrix elements which is not reliable so far, namely we cannot handle
the non-perturbative QCD effects well, the deviation between the two
$\delta$ values is reasonable.

In conclusion, we propose that the weak CP phase in the CKM matrix is
the geometric
phase for an O(3) rotation in the flavor space, and determined by the three
mixing angles according to Eq.(4), its value is consistent
with that extracted from the measurement of $\epsilon$ in K-system.

Meanwhile, to make the three rotation angles enclose a solid angle,
the following constraint must be satisfied
$$\theta_i+\theta_j\geq \theta_k,\;\;\; (i\neq j\neq k \;{\rm and}\; i,j,k
=1,2,3).$$
It would provide a criterion for judging our postulation.

It is also interesting to make a comparison with the four-generation case.
There a rotation in the flavor space should be four-dimensional,
correspondingly, six rotation angles
and three independent phases should be present. That is obviously
coincide with the parametrization in four dimensional CKM matrix where
the number of rotation angles is $N(N-1)/2=6$ while the number of phases is
$(N-1)(N-2)/2=3$.

Anyway, the relation shown in Eq.(4) postulated by us can give some strict
limits on the free parameters presented in KM matrix, just so, it can be
further put to the more precise tests in the future. At least, Eq.(4) can
be taken as a good parameterization form for the weak CP-violation phase.

\vspace{0.5cm}

\noindent {\bf Acknowledgment}: This work is partly supported by National
Natural Science Foundation of China. We would like to thank Dr. Z.Z. Xing
for helpful comments.

\end{document}